\documentclass[a4paper,11pt]{article}
\usepackage{pos}
\usepackage{amsmath}
\usepackage{graphicx}

\usepackage{cleveref}
\crefname{equation}{Eq.}{Eqs.}
\Crefname{equation}{Equation}{Equations}
\crefname{section}{Section}{Sections}
\Crefname{section}{Section}{Sections}
\crefname{subsection}{Section}{Sections}
\Crefname{subsection}{Section}{Sections}

\newcommand{\eqnDiag}[1]{ \vcenter{\hbox{#1}} }

\begin{document}

\title{Compact Syzygies for Feynman Integrals from Landau Singularities}

\author[a]{Federico Coro}
\author[a]{Pavel P.~Novichkov}
\author*[a]{Ben Page}
\author[a]{Qian Song}

\affiliation[a]{Department of Physics, University of Ghent,
  \\
Krijgslaan 281, S9, B-9000 Ghent, Belgium}

\emailAdd{ben.page@ugent.be}
\emailAdd{federico.coro@ugent.be}
\emailAdd{pavel.novichkov@ugent.be}
\emailAdd{qian.song@ugent.be}

\abstract{
We discuss a recent proposal for constructing ``syzygy solutions'', which play a
crucial role in integration-by-parts reductions for multi-loop scattering
amplitudes.
We highlight a relationship between syzygies and the leading Landau
singularities of Feynman diagrams and discuss how this implies that integral
relations are controlled by infrared singularities.
We use this insight to systematically construct highly compact, determinantal
syzygy solutions, providing an explicit, simple example at one loop.
We demonstrate the power of this approach by applying it to a two-loop pentabox
family relevant for the NNLO corrections to $pp \rightarrow t\overline{t}H$ at
the LHC.
}

\FullConference{Loops and Legs in Quantum Field Theory (LL2026)\\
12-17, April, 2026\\
Bayreuth, Germany\\}

\maketitle

\section{Introduction}
\label{sec:introduction}

Computations of perturbative scattering amplitudes are of great importance for the research program at particle colliders and gravitational-wave experiments.
At multi-loop order, these computations yield gigantic integrands: a single Feynman diagram can easily exceed hundreds of megabytes in size.
To integrate such formidable expressions, it is standard practice to reduce the vast number of arising Feynman integrals down to a small basis of master integrals.
This reduction is performed by employing integration-by-parts (IBP) identities~\cite{Chetyrkin:1981qh} based on the vanishing of total derivatives in dimensional regularization.
The IBP identities can be interpreted as a massive system of linear equations among the integrals.
Solving this system systematically leads to the well-known Laporta algorithm~\cite{Laporta:2000dsw}.
Today, a variety of active public implementations of this approach exist~\cite{
  Smirnov:2008iw,Smirnov:2025prc, 
  Maierhofer:2017gsa,Klappert:2020nbg, 
  Guan:2024byi, 
  Wu:2023upw,Wu:2025aeg, 
  Peraro:2019svx, 
  Lee:2012cn,Lee:2013mka}. 

Despite many algorithmic advances, IBP reduction of Feynman integrals remains a
bottleneck of frontier perturbative calculations.
One difficulty is severe expression swell. To overcome this, modern techniques
often reconstruct the final reduction coefficients from finite-field samples,
evaluating the system modulo a prime
number~\cite{vonManteuffel:2014ixa,Peraro:2016wsq}.
Complementary strategies focus on directly decreasing the size of the linear system before sampling.
This can be achieved by utilizing syzygies to avoid doubled propagators~\cite{Gluza:2010ws,Larsen:2015ped,Ita:2015tya}, or by shifting the system into block triangular form~\cite{Guan:2019bcx}.
Alternatively, one can bypass traditional IBP frameworks altogether through the geometric machinery of intersection theory~\cite{Mastrolia:2018uzb}.

In this contribution, we report on a recent approach~\cite{Page:2025gso,Coro:2025kha} that organizes integral relations according to the infrared (IR) singularity structure of the underlying Feynman integrals.
In \cref{sec:theory}, we discuss how syzygies are intimately connected to so-called leading Landau singularities and use this connection to derive compact determinantal syzygy solutions.
In \cref{sec:examples}, we illustrate our method with a simple one-loop example, as well as a more complex two-loop pentabox integral family computation relevant for the \(pp \rightarrow t\overline{t}H\) process.
We summarize in \cref{sec:conclusions}.

\section{Syzygies from Landau singularities}
\label{sec:theory}

\subsection{Syzygies in Baikov representation}

We find it convenient to work with the Baikov representation~\cite{Baikov:1996rk} of Feynman integrals.
For concreteness, we will focus on one- and two-loop Feynman diagrams~\(\Gamma\) with five or more external legs.
We split the loop momenta into four- and \((-2 \epsilon)\)-dimensional parts as \(\ell_a + \ell_a^\epsilon\); the external momenta are taken to be strictly four-dimensional.
Denoting \(\mu_{ab} = \ell_a^\epsilon \cdot \ell_b^\epsilon\), we can write the integral (up to an irrelevant prefactor) as~\cite{Larsen:2015ped}
\begin{equation}
  \label{eq:BaikovRepresentation}
  I_\Gamma \left[ \mathcal{N} \right] =
  \int \prod_{1 \leq a \leq b \leq L} \mathrm{d} \mu_{ab} \prod_{1 \leq a \leq L} \mathrm{d}^4 \ell_a \, B^{-(L+1)/2 - \epsilon} \,
  \frac{\mathcal{N} (\ell_a, \mu_{ab})}{\prod_{e \in \mathrm{props}(\Gamma)} z_e}.
\end{equation}
Here \(L = 1\) or 2 is the number of loops, $z_e$ denote the denominators
corresponding to the propagators of the diagram, \(\mathcal{N}\) is the integral numerator, and $B = \det (\mu_{ab})$ is the normalized Baikov polynomial.

Instead of the loop-momentum variables \(\left\{ \ell_a, \mu_{ab} \right\}\), one could express the integral~\eqref{eq:BaikovRepresentation} in terms of the
Baikov variables formed by denominators \(z_e\) and irreducible scalar products (ISPs) \(z_i\).
These two sets of variables are related by a one-to-one quadratic map with a constant Jacobian (see e.g.\ Ref.~\cite{Larsen:2015ped}), and thus can be used interchangeably.
We prefer to work with the loop-momentum variables as they lead to compact, physically transparent expressions.

One can systematically avoid generating IBP relations with unwanted doubled propagators by constructing identities from the so-called ``syzygy'' equation~\cite{Gluza:2010ws,Larsen:2015ped,Ita:2015tya,Page:2025gso}:
\begin{equation}
  \label{eq:SyzygyEquation}
  a_0 B + \sum_{e \in \mathrm{props}(\Gamma)} \tilde{a}_e z_e B + \sum_{i \in \mathrm{ISPs}(\Gamma)} a_i \partial_i B + \sum_{e \in \mathrm{props}(\Gamma)} \overline{a}_e z_e \partial_e B = 0.
\end{equation}
Here, \(\partial_e\) and \(\partial_i\) denote partial derivatives with respect to the Baikov variables, and one solves the equation for \(\vec{a} = (a_0, \tilde{a}_e, a_i, \overline{a}_e)\) --- a tuple of unknown polynomials in the loop-momentum (or Baikov) variables whose coefficients are rational in the external kinematics and masses.
Each solution gives rise to an integral relation free of doubled denominators:
\begin{equation}
  0 =
  \int \prod_{1 \leq a \leq b \leq L} \mathrm{d} \mu_{ab} \prod_{1 \leq a \leq L} \mathrm{d}^4 \ell_a \, B^{-(L+1)/2 - \epsilon} \,
  \frac{S_\Gamma (\vec{a})}{\prod_{e \in \mathrm{props}(\Gamma)} z_e},
\end{equation}
where
\begin{equation}
  \label{eq:SurfaceTerm}
  S_\Gamma (\vec{a}) =
  a_0
  + \sum_{e \in \mathrm{props}(\Gamma)} \tilde{a}_e z_e
  + \frac{1}{(L+1)/2 + \epsilon} \left[
    \sum_{i \in \mathrm{ISPs}(\Gamma)} \partial_i a_i
    +  \sum_{e \in \mathrm{props}(\Gamma)} z_e \partial_e \overline{a}_e
  \right]
\end{equation}
is the associated so-called surface term.

One can easily construct some solutions of \cref{eq:SyzygyEquation} by dotting
any antisymmetric (polynomial) matrix into the vector \((B, z_e B, \partial_i B, z_e \partial_e B)\).
Such solutions are called ``principal syzygies''.
However, in many cases additional solutions exist; finding them is a highly non-trivial problem.

\subsection{A syzygy by inspection}
\label{sec:4dSyzygy}

Let us consider an elementary construction of a non-principal syzygy,
reproducing a syzygy first found in Ref.~\cite{Coro:2025kha}.
We will exploit the algebraic structure of the two-loop Baikov polynomial, which
alongside its derivatives can be written in terms of only three objects \(\mu_{ab}\):
\begin{equation}
  \label{eq:4dRelations}
  \partial_i B = \mu_{22} \partial_i \mu_{11} + \mu_{11} \partial_i \mu_{22} - 2 \mu_{12} \partial_i \mu_{12}, \qquad B = \mu_{11} \mu_{22} - \mu_{12}^2.
\end{equation}
That is, the Baikov polynomial and $\mu_{ab}$ derivatives are linearly dependent.
If the number of ISPs is three or more, we can assemble \cref{eq:4dRelations}
into a matrix $\mathcal{F}$ that is guaranteed to be rank-deficient, e.g.
\begin{equation}
  \mathcal{F} = \begin{bmatrix}
    B & \frac{1}{2}\mu_{11} & \frac{1}{2}\mu_{22} & \frac{1}{2} \mu_{12} \\
    \partial_1 B & \partial_1 \mu_{11} & \partial_1 \mu_{22} & \partial_1 \mu_{12} \\
    \partial_2 B & \partial_2 \mu_{11} & \partial_2 \mu_{22} & \partial_2 \mu_{12} \\
    \partial_3 B & \partial_3 \mu_{11} & \partial_3 \mu_{22} & \partial_3 \mu_{12}
  \end{bmatrix},
  \qquad \mathcal{F} \begin{bmatrix} -1 \\ \mu_{22} \\ \mu_{11} \\ -2 \mu_{12} \end{bmatrix} = 0 \quad \Rightarrow \quad \det[\mathcal{F}] = 0.
  \label{eq:4dCertificateMatrix}
\end{equation}
Evaluating the determinant of $\mathcal{F}$ via a Laplace expansion along its first column naturally generates a compact syzygy solution:
\begin{equation}
  0 = \text{minor}_{1,1}(\mathcal{F}) \times B + \sum_{i=1}^3 (-1)^i \text{minor}_{1, i+1}(\mathcal{F}) \times (\partial_i B).
\end{equation}
Our main result, described next, is a systematic generalization of this construction.

\subsection{Critical syzygies and Landau singularities}
\label{sec:CriticalSyzygies}

It was observed in Ref.~\cite{Page:2025gso} that the \(a_0\) term of
\cref{eq:SyzygyEquation} plays a distinguished role in building surface terms
for IBP relations: indeed, \cref{eq:SurfaceTerm} implies that  the \(a_0\) term \emph{is} the surface term in the large-\(\epsilon\) limit on the maximal cut \(z_e = 0\), \(e \in \text{props}(\Gamma)\):
\begin{equation}
  a_0 = \lim_{\epsilon \to \infty} \left. S_\Gamma (\vec{a}) \right|_{z_e = 0}.
\end{equation}
Consequently, two distinct syzygy solutions $\vec{a}^{[1]}$ and $\vec{a}^{[2]}$ with the same $a_0$ on the cut become equivalent in this limit as they yield the same integral relation:
\begin{equation}
  \label{eq:CriticalEquivalence}
  \vec{a}^{[1]} \sim \vec{a}^{[2]} \; \leftrightarrow \; \left. a_0^{[1]} \right|_{z_e = 0} = \left. a_0^{[2]} \right|_{z_e = 0} \; \leftrightarrow \; \lim_{\epsilon \rightarrow \infty} \left.\left[  S_\Gamma(\vec{a}^{[1]}) - S_\Gamma(\vec{a}^{[2]})  \right]\right|_{z_e = 0} = 0.
\end{equation}
We say that \(\vec{a}^{[1]}\) and \(\vec{a}^{[2]}\) define the same \emph{critical syzygy}.
One of the main results of Ref.~\cite{Page:2025gso} is that in many cases%
\footnote{More precisely, this holds if \(B\) has isolated critical points and the ideal \(\left\langle z_e B, \partial_i B, z_e \partial_e B \right\rangle : \left\langle B \right\rangle^\infty\) has saturation index 1.}
critical syzygies \(\text{CSyz}(\Gamma)\) generate all relevant integral relations also away from the large-\(\epsilon\) limit: in other words, to construct all surface terms it suffices to take a single representative from each equivalence class defined by \cref{eq:CriticalEquivalence}.

Motivated by this observation, we focus on constructing the \(a_0\) part of the
syzygy. As can be seen from \cref{eq:SyzygyEquation}, the \(a_0\) term must
vanish on
\begin{equation}
  U_{\text{syz}}^\Gamma = \big\{ z_e B = \partial_i B = z_e \partial_e B = 0 \; : \; e \in \mathrm{props}(\Gamma), \; i \in \mathrm{ISPs}(\Gamma) \big\}
\end{equation}
(that is, whenever all but the first term vanish), except for the subset \(U_{\text{sing}}^\Gamma\) where \(B\) vanishes as well:
\begin{equation}
  U_{\text{sing}}^\Gamma = U_{\text{syz}}^\Gamma \cap \left\{ B = 0 \right\}.
\end{equation}
For principal syzygy solutions, \(a_0\) vanishes everywhere on \(U_{\text{syz}}^\Gamma\) including \(U_{\text{sing}}^\Gamma\).
Hence, we can construct interesting \(a_0\)'s by forcing their non-vanishing on some part of \(U_{\text{sing}}^\Gamma\).

It turns out that this can be achieved with a determinantal procedure.
Suppose that the part of \(U_{\text{sing}}^\Gamma\) we wish to excise is the vanishing locus of a set of polynomials \(\vec{b}\), and we managed to express the polynomials describing \(U_{\text{syz}}^\Gamma\) in terms of \(\vec{b}\) via a matrix equation:
\begin{equation}
  U_{\vec{b}} = \left\{ \vec{b} = 0 \right\} \subset U_{\text{sing}}^\Gamma, \qquad
  \begin{bmatrix}
    \partial_i B \\ z_e \partial_e B
  \end{bmatrix}
  = M \vec{b}.
\end{equation}
Then, \(M \vec{b}\) vanishes on \(U_{\text{syz}}^\Gamma\), so that \(M\) drops rank on \(U_{\text{syz}}^\Gamma \setminus U_{\vec{b}}\), and we can construct \(a_0\)'s as maximal minors of \(M\).

The syzygy of \cref{sec:4dSyzygy} gives an example of this construction.
Here, \(\vec{b} = (\mu_{22}, \mu_{11}, -2 \mu_{12})\) and \(M\) is a submatrix of \(\mathcal{F}\) without its first row and column, so that \((\partial_i B) = M \vec{b}\).
This example also shows that our determinantal construction of \(a_0\) extends to the full syzygy solution, if we additionally express \(B\) in terms of \(\vec{b}\) and rearrange the matrix equation in a form similar to \cref{eq:4dCertificateMatrix} with a ``certificate matrix'' \(\mathcal{F}\):
\begin{equation}
  \mathcal{F} =
  \begin{bmatrix}
    B & \vec{c}^T \\
    \partial_i B & \mathcal{C}^{[i]} \\
    z_e \partial_e B & \mathcal{C}^{[e]} \\
    \vec{0} & \mathcal{R}
  \end{bmatrix},
  \qquad \mathcal{F} \begin{bmatrix} -1 \\ \vec{b} \end{bmatrix} = 0,
\end{equation}
where \(\mathcal{R}\) is a matrix describing additional relations (non-principal syzygies) among \(\vec{b}\), if any.

How does one systematically construct \(U_{\vec{b}} \subset U_{\text{sing}}^\Gamma\)?
In Ref.~\cite{Coro:2025kha} we argued that a useful answer is provided by the singularity structure of the underlying Feynman integral.
Namely, \(U_{\text{sing}}^\Gamma\) decomposes as the union of leading Landau singularities~\cite{Landau:1959fi,Caron-Huot:2024brh,Correia:2025wtb} of \(\Gamma\) and its subdiagrams:
\begin{equation}
  U_{\text{sing}}^\Gamma = \bigcup_{\Gamma' \subseteq \Gamma} U_{\text{Landau}}^{\Gamma'}, \quad
  U_{\text{Landau}}^{\Gamma'} =
  \big\{
    B = \partial_i B = z_e = 0 \; : \; e \in \text{props}(\Gamma'), \; i \in \text{ISPs}(\Gamma')
  \big\}.
\end{equation}
Physically, leading Landau singularities describe IR singularities that are compatible with the cut.
A leading singularity ``extends'' off the cut to a union of minimal IR singularities intersecting it.
We take this extension, called \emph{prelocalization} in Ref.~\cite{Coro:2025kha}, to be the excised surface \(U_{\vec{b}}\) in our determinantal construction.
(Such extension is necessary to ensure completeness of the resulting set of syzygies.)

In this way, each leading Landau singularity of \(\Gamma\) gives a set of compact determinantal syzygies whose \(a_0\) term does not vanish on this singularity.
Combining results for all leading singularities, we are guaranteed to find the full set of critical syzygies if a certain technical assumption holds~\cite{Coro:2025kha}.
More formally, the critical syzygy module can be written as a module sum distributed across the constituent irreducible components of the maximal cut Landau locus:
\begin{equation}
  \label{eq:CSyzDecomp}
  U_{\text{Landau}}^\Gamma = \bigcup_j U_j^\Gamma \quad \Rightarrow \quad
  \text{CSyz}(\Gamma) = \sum_j \text{FittSyz}(U^{\Gamma}_j),
\end{equation}
where each $\text{FittSyz}(U^{\Gamma}_j)$ is generated by minors of a matrix $\mathcal{F}_j$ which certifies the singularity.

\section{Examples and Applications}
\label{sec:examples}

\subsection{The One-Loop Triangle}
To illustrate our approach in a concrete, pedagogical, example, we
consider a simple one-loop diagram, the so-called ``one-mass triangle''.
We treat it as a subtopology of the pentagon diagram with massless outgoing external momenta $p_i$, $i=1, \dots, 5$, subject to momentum conservation:
\begin{equation}
  \eqnDiag{\includegraphics[scale=0.7]{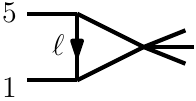}}
  \qquad
  p_i^2 = 0, \quad
  \sum_{i=1}^5 p_i = 0. 
\end{equation}
At one loop, we suppress the loop-momentum indices, writing simply \(\ell^\mu \equiv \ell_1^\mu\), \(\mu^2 \equiv \mu_{11}\).

We define the Baikov variables in terms of partial sums of the external momenta \(q_k\):
\begin{equation}
  q_k = \sum_{j=1}^k p_j, \qquad
  z_k = (\ell - q_k)^2, \quad
  k = 1, \dots, 5.
\end{equation}
Here, \(z_1, z_4, z_5\) correspond to the propagators, while \(z_2, z_3\) are the ISPs.
We also introduce a dual vector basis $v_i$, \(i = 1, \dots, 4\), satisfying the orthogonality condition \(v_i \cdot q_j = \delta_{ij}\).

A direct computation shows that the maximal-cut Landau locus has only one component, namely the soft singularity:
\begin{equation}
  U_{\text{Landau}}^{\text{tri}} = \left\{ \ell^\mu = \mu^2 = 0 \right\}.
\end{equation}
(We perform explicit computations with \textsc{Singular}~\cite{DGPS}, working in loop-momentum variables with numerical external kinematics.)
Off the cut, it extends to the union of two collinear singularities:
\begin{equation}
  \begin{aligned}
\eqnDiag{\includegraphics[scale=0.6]{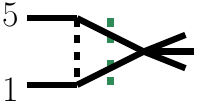}}
    \; \rightarrow \;
    \eqnDiag{\includegraphics[scale=0.5]{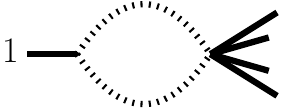}}
    \;\; &\cup \;\;
    \eqnDiag{\includegraphics[scale=0.5]{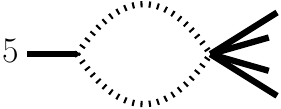}}
    \\
    \sim \left\{ \ell_\perp^\mu = z_5 = z_1 = 0\right\} &\cup \left\{ \ell_\perp^\mu = z_5 = z_4 = 0 \right\}
    = \left\{ \ell_\perp^\mu = z_5 =  z_1 z_4 = 0 \right\},
\end{aligned}
\label{eq:TriangleLandauOrganization}
\end{equation}
where \(\ell_\perp^\mu\) is the projection of the loop momentum onto the
two-dimensional space orthogonal to \(p_1, p_5\), and should be regarded as a
two-component object.
In
\cref{eq:TriangleLandauOrganization}, the green dashed lines decorate cut
propagators and the black dashed (dotted) lines denote propagators where the line
momentum is soft (collinear).

We construct the certificate matrix by inspection, expressing \(B = \mu^2\) and its derivatives in terms of \(\vec{b} = (\ell_\perp^\mu, z_5, z_1 z_4)\):
\begin{equation}
  \begin{bmatrix}
    B & - \ell_\perp^\mu & s_{15} - z_4 - z_1 + z_5 & 1 \\
    \partial_2 B & v_{2 \perp}^\mu & 0 & 0 \\
    \partial_3 B & v_{3 \perp}^\mu & 0 & 0 \\
    z_4 \partial_4 B & z_4 v_{4 \perp}^\mu & -z_4 & 1 \\
    z_5 \partial_5 B & 0_\perp^\mu & s_{15} \partial_5 B & 0 \\
    z_1 \partial_1 B & z_1 v_{1 \perp}^\mu & -z_1 & 1
  \end{bmatrix}
  \;
  \begin{bmatrix} 
    -1 \\
    \ell_\perp^\mu \\
    \frac{1}{s_{15}} z_5 \\
    \frac{1}{s_{15}} z_1 z_4
  \end{bmatrix}
  = 0,
\end{equation}
where \(s_{15} = (p_1 + p_5)^2\). The certificate matrix is a $6 \times 5$
matrix, and syzygy solutions can now be easily computed from its maximal minors.

\subsection{Two-Loop Applications}
To demonstrate the power of our approach, let us consider
a more complex application beyond one loop.
Namely, we examine a two-loop
double-box integral family relevant for massive top-quark and Higgs
phenomenology, specifically $pp \rightarrow t\overline{t}H$:
\begin{equation}
  \eqnDiag{\includegraphics[scale=0.45]{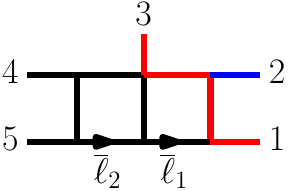}} \qquad p_4^2 = p_5^2 = 0, \,\, p_1^2 = p_3^2 = m_t^2, \, \, p_2^2 = q^2,
  \,\,
  \sum_{i=1}^5 p_i = 0.
\end{equation}
For this topology, the maximal-cut Landau locus breaks up into three components:
\begin{equation}
    U^{\text{db}}_{\text{Landau}} = U^{\text{db}}_0 \cup U^{\text{db}}_1 \cup U^{\text{db}}_2.
\end{equation}
The first component corresponds to the maximal-cut four-dimensional locus
\begin{equation}
          U_0^{\text{db}} = \left\{ \mu_{11} = \mu_{22} = \mu_{12} = z_e = 0 \; : \; e \in \text{props}(\text{db}) \right\}.
\end{equation}
The remaining components correspond to configurations where internal particles
become either soft or collinear:
\begin{equation}
  U_{1}^{\text{db}} \,\, \sim \,\,  \eqnDiag{\includegraphics[scale=0.45]{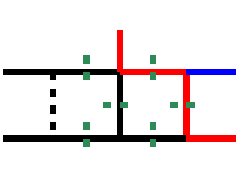}} , \qquad
  U_{2}^{\text{db}} \,\, \sim \,\,  \eqnDiag{\includegraphics[scale=0.45]{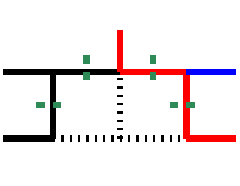}}.
\end{equation}
We then apply our determinantal procedure to construct the syzygy solutions;
the matrix $\mathcal{F}$ of \cref{eq:4dCertificateMatrix}
can be seen to be (a submatrix of) the certificate matrix associated with the
$U^{\text{db}}_0$ singularity and the certificate matrices associated to the
other singularities can be constructed with the recipe of \cref{sec:CriticalSyzygies}.

In this way, in Ref.~\cite{Coro:2025kha} we
constructed the complete set of critical syzygy relations for the five-point
topologies of the $pp \rightarrow t \overline{t}H$ pentabox family, that is, the
pentabox along with its associated pentatriangle, box-triangle, and pentabubble
subsectors:
\begin{equation}
  \eqnDiag{\includegraphics[scale=0.35]{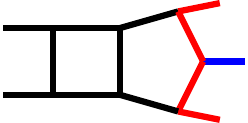}} \quad
  \eqnDiag{\includegraphics[scale=0.35]{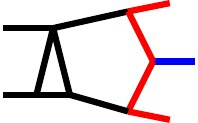}} \quad
  \eqnDiag{\includegraphics[scale=0.35]{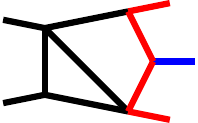}} \quad
  \eqnDiag{\includegraphics[scale=0.35]{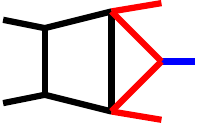}} \quad
  \eqnDiag{\includegraphics[scale=0.35]{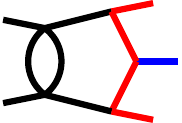}}.
\end{equation}
Crucially, many of the required off-shell extensions remain identical across
distinct topologies. For instance, the external soft configurations for the
pentabox, double-box and box-triangle sectors extend to the exact same
pair of collinear singularities, i.e.,
\begin{equation}
  \eqnDiag{\includegraphics[scale=0.35]{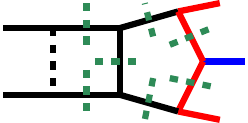}},
  \quad
  \eqnDiag{\includegraphics[scale=0.35]{graphics/DoubleBoxExternalSoft.pdf}},
  \quad \,\,
  \eqnDiag{\includegraphics[scale=0.35]{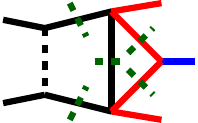}}
  \qquad
  \to
  \qquad
  \eqnDiag{\includegraphics[scale=0.3]{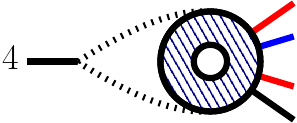}}
  \; \cup \;
  \eqnDiag{\includegraphics[scale=0.3]{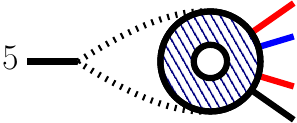}}.
\end{equation}
This observation means that it is possible to construct the $\mathcal{F}_j$
matrices in such a way that we can recycle the calculation between the different
five-point integrals that share the same underlying IR singularities.
Thus, we see that, in our construction, the universality of IR
divergences is inherited by the syzygy solutions.

\section{Conclusions}
\label{sec:conclusions}

In these proceedings, we discussed a novel method for the construction of
compact syzygy solutions in a determinantal fashion.
We worked in the ``critical syzygy'' formalism, which exploits structures
arising in the large-$\epsilon$ limit of dimensional regularization to single
out a small subset of the full collection of syzygies that are required for
multi-loop integral reduction at any value of $\epsilon$.
We used this perspective to highlight a connection between the Landau equations
and the syzygy equation and argued that Feynman integral relations are
controlled by the infrared singularities that underlie the associated diagrams.
We exploited this connection in an algebraic fashion to yield
determinantal expressions for syzygies.
As these formulae are exceptionally compact, we are optimistic
that they will lead to highly efficient, automatic approaches for Feynman
integral reduction.

\acknowledgments{
  The work of Federico Coro, Pavel Novichkov and Qian Song was supported by the European Research Council (ERC) under the European Union’s Horizon Europe research and innovation program grant agreement 101078449 (ERC Starting Grant MultiScaleAmp).
  Views and opinions expressed are however those of the authors only and do not necessarily reflect those of the European Union or the European Research Council Executive Agency.
  Neither the European Union nor the granting authority can be held responsible for them.
}

\bibliographystyle{JHEP}
\bibliography{pos}

\end{document}